\documentclass[onecolumn,showpacs,preprintnumbers,amsmath,amssymb]{revtex4-1}

\usepackage{graphicx}
\usepackage{amssymb}
\usepackage{hyperref}

\renewcommand
\baselinestretch{1.5}

\begin{document}

\title{Low temperature thermodynamic properties of Ce compounds\\ tuned at $T_{ord}=0$}

\author{Julian G. Streinz-Sereni$^*$}

\address{Low Temperature Division, CAB-CNEA, CONICET, 8400 S.C. de Bariloche, Argentina}

\date{\today}

\begin{abstract}
{Based on specific heat ($C_m$) and entropy evaluation, different
Ce magnetic phase diagrams can be recognized: I) with the entropy
of the ordered phase ($S_{MO}$) decreasing with their order
temperature ($T_{ord}$), which are the only candidates for quantum
critical behavior since $S_{MO}\to0$ as $T_{ord}\to0$. II) with
phase boundaries ending at a $T>0$ critical point because their
$C_m(T_{ord})$ jumps ($\Delta C_m$) do not decrease sufficiently
with $T_{ord}$ producing a $S_{MO}$ bottleneck, and III) those
showing a transference of degrees of freedom to a non-magnetic
component, with their $\Delta C_m$ vanishing at $T>>0$.

IV) There is also a group of Ce heavy fermions which do not order
magnetically down to $T\cong 0$. These compounds are at the top of
the $\lim_{T\to 0}
\partial S_m /\partial T$= $C_m/T$ values because they  collect
very high density of low energy excitations. From the analysis of
$C_m(T)/T$ and $S_m(T)$ results performed on selected Ce
ternaries, a quantitative determination of an upper limit for the
density of excitations is obtained, excluding any evidence of
$C_m(T)/T$ divergency at $T\to 0$ in agreement with thermodynamic
laws.

\vspace{0.3cm}

$^*$ E-mail-address of corresponding author:
jsereni@cab.cnea.gov.ar}

\end{abstract}

\renewcommand
\baselinestretch{1}

\pacs{75.20.Hr, 71.27.+a, 75.30.Kz, 75.10.-b}

\maketitle

\section{Introduction.}

\begin{figure}
\begin{center}
\includegraphics[angle=0, width=0.6 \textwidth] {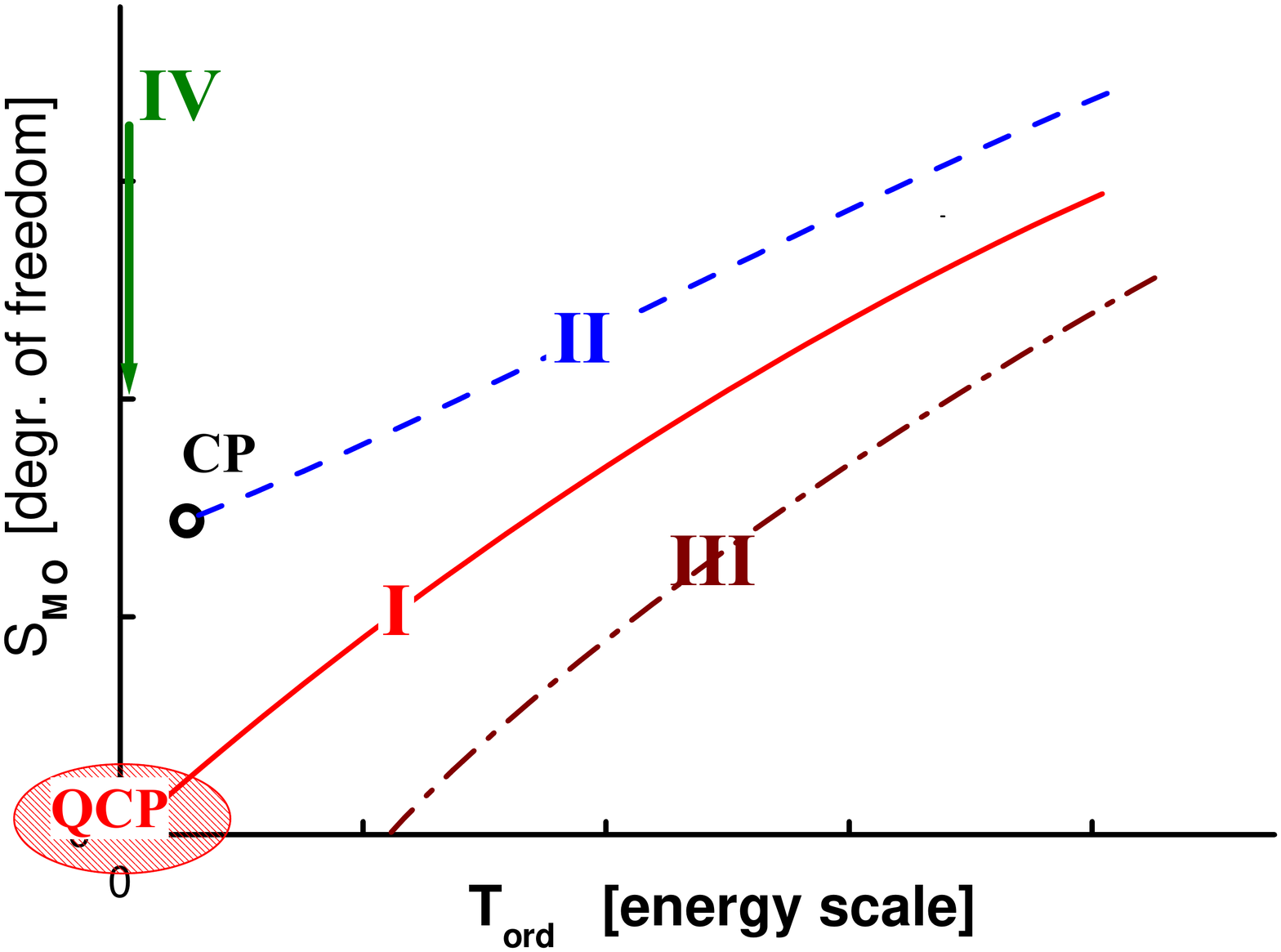}
\end{center}
\caption{Schematic comparison of three types of magnetic phase
diagrams for the entropy of the ordered phase ($S_{MO}$) collected
up to the transition at $T_{ord}$, after ref. \cite{PhilMag}. The
arrow label as {\bf 'IV'} represents Ce compounds which do not
show magnetic order down to $T\cong 0$ analyzed in this work.}
\label{F1}
\end{figure}

Among the outstanding subjects of current investigations on
strongly correlated electron systems are those related with
quantum criticality (QC) \cite{HvL}. Despite of the large amount
of new intermetallic compounds claimed to be candidates to reach a
quantum critical point (QCP) at $T=0$, only a few of them were
proved to reach that regime. Simple thermodynamic principles
clearly establish the conditions for such scenario, which can be
tested through the entropy (i.e the degrees of freedom) collected
into the ordered phase $S_{MO}$ up to the ordering temperature
$T_{ord}$ according to the $S_{MO}\to 0$ as $T_{ord}\to 0$
criterion \cite{PhilMag}. This and alternative types of phase
diagrams are schematically summarized in Fig.~\ref{F1}

The condition for $S_{MO}\to 0$ as $T_{ord}\to 0$ is fulfilled for
systems represented by curve {\bf 'I'}. The alternative types of
magnetic phase diagrams can be recognized from the analysis of the
specific heat jump at $T_{ord}$ ($\Delta C_{MO}$) \cite{PhilMag}.
Those ending at a finite temperature critical point because their
$\Delta C_{MO}$ do not decrease sufficiently as $T_{ord}$
decreases with a consequent bottleneck of $S_{MO}$ at low
temperature, label as {\bf 'II'} in Fig.~\ref{F1}. Another group
shows a transference of degrees of freedom from the MO phase to a
non-magnetic component with their $\Delta C_m$ vanishing at finite
temperature, represented by curve {\bf 'III'} in Fig.~\ref{F1}.

Despite of the clear thermodynamic constraint (c.f. $S_{MO} \to 0$
when $T_{ord} \to 0$) for reaching a QCP the phase diagrams of
real systems exhibit intrinsic differences with theoretical
predictions. For example, their 2nd order magnetic phase
boundaries at $T_{ord}(x)$ driven by a 'non-thermal' control
parameter \cite{TVojta} like chemical potential, do not decrease
according to the expected negative curvature down to $T\to 0$
because they deviate from that monotonous behavior at $T^{CR}
\approx 2.5$\,K \cite{PhilMag}. This experimental observation
invalidates misleading extrapolations of $T_{ord}(x) \to 0$
preformed according to conventional criteria. A pre-critical
region is identified below $T^{CR}$ where the nature of the
magnetic phase boundary undergoes significant modifications, like
e.g. to become of 3rd order with a linear $T_{ord}(x)$ dependence.
That 3rd order transition, identified from a jump in $\partial C_m
/\partial T$ \cite{Pippard}, is suppressed by moderate magnetic
field without changing $T_{ord}$. Also an increasing remnant
entropy (up to 0.4\,R$Ln$2) is observed in this critical region as
$T\to 0$ \cite{anivHvL}.

\begin{figure}[t]%
\includegraphics*[width=0.6 \linewidth] {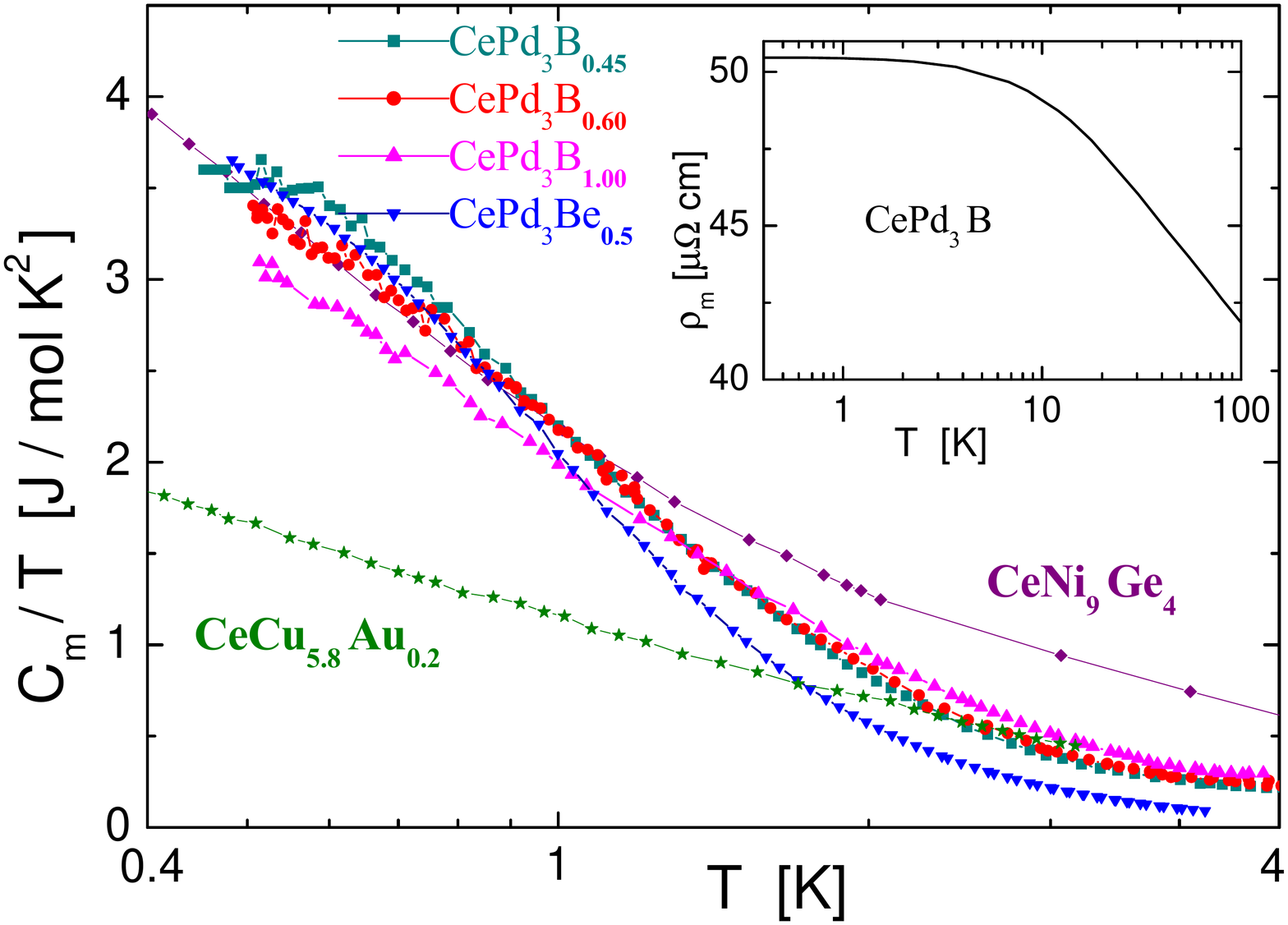}
\caption{(Color online) Low temperature specific heat of the
heaviest fermions reported among Ce-lattice compounds in a log T
representation. The well known heavy fermion (HF)
CeCu$_{5.8}$Au$_{0.2}$ \cite{HvL1999} is included for comparison.
Inset: low temperature magnetic contribution to the electrical
resistivity of CePd$_3$B after \cite{Gladys} in a logarithmic
temperature scale.} \label{F2}
\end{figure}

In this work we analyze a fourth group of Ce compounds (see arrow
label as {\bf 'IV'} in Fig.~\ref{F1}) which do not order
magnetically because they behave as tuned at $T_{ord}=0$. These
systems collect the highest density of low energy excitations
reflected in huge values of $C_m/T$ at $T\to 0$. They represent an
empirical upper limit for $\lim_{T\to 0} \partial S_m/\partial T$=
$C_m/T$ because the systems showing magnetic order condense their
magnetic degrees of freedom into the ordered phase between
$T_{ord} > T > 0$.

\section{Experimental results}

A record high $C_m/T$ value at low temperature was claimed for the
compound CePd$_3$B more than two decades ago \cite{CePd3B}.
Despite of the lack of magnetic order down to $T=0.5K$ there were,
however, some indications for the onset of short range magnetic
correlations from a broad maximum around $T=0.7$\,K. A new series
of samples were produced improving the annealing process which is
reflected in a smoother $C_m(T)/T$ dependence around that
temperature.

In Fig.~\ref{F2} we present the experimental results on the new
CePd$_3$B$_x$ series with $x = 0.45, 0.60, 1$ and
CePd$_3$Be$_{0.5}$ down to $T\approx 0.5$\,K in a logarithmic
temperature scale. The magnetic contribution was obtained after
subtracting the phonon component extracted from LaPd$_3$B
\cite{Gladys}. For comparison, also CeNi$_9$Ge$_4$
\cite{ScheidtGe4} and CeCu$_{5.8}$Au$_{0.2}$ \cite{HvL1999}
compounds are included. The former shows the highest
$C_m/T\mid_{\lim T\to 0}$ value reported at present among
Ce-lattice intermetallics, whereas the later is the prototype of
non-Fermi-liquids (NFL) accessing to QC regime with a logarithmic:
$C_m/T \propto -Ln(T/T_0)$ dependence. CeNi$_9$Ge$_4$ was measured
down to $\approx 50$\,mK, where it reaches a record value of
$C_m/T = 5.5$\,J/molK$^2$. Such an extremely high value is
associated to the contribution of the ground state and first
excited crystal field (ECF) levels with a splitting comparable to
the Kondo temperature $T_K = 10$\,K \cite{ScheidtGe4} which makes
this system to be considered as an effective $N_{eff}>2$
degenerated ground state (GS). Nevertheless, in Fig.~\ref{F2} it
can be seen that CePd$_3$B$_x$ compounds reach similar $C_m/T$
values for $0.5 \leq T \leq 1$\,K despite of their $N=2$ GS
character. Such a coincidence can be explained by the irrelevant
Kondo effect in these cubic ternary compounds. From their magnetic
properties one may distinguish between an effective $N_{eff}>2$ GS
and an actual $N=4$. In the former two Kramer doublets with
different giromagnetic factors ($g_i$) overlap in energy because
the ECF splitting ($\Delta_I$) compares with their level
broadening (i.e. $\Delta_I/T_K\approx 1$). On the contrary, in the
$N=4$ case Kondo effect may be even absent like in BCC Ce-binary
compounds \cite{Handb}.

\begin{figure}[t]%
\includegraphics*[width=0.6\linewidth]{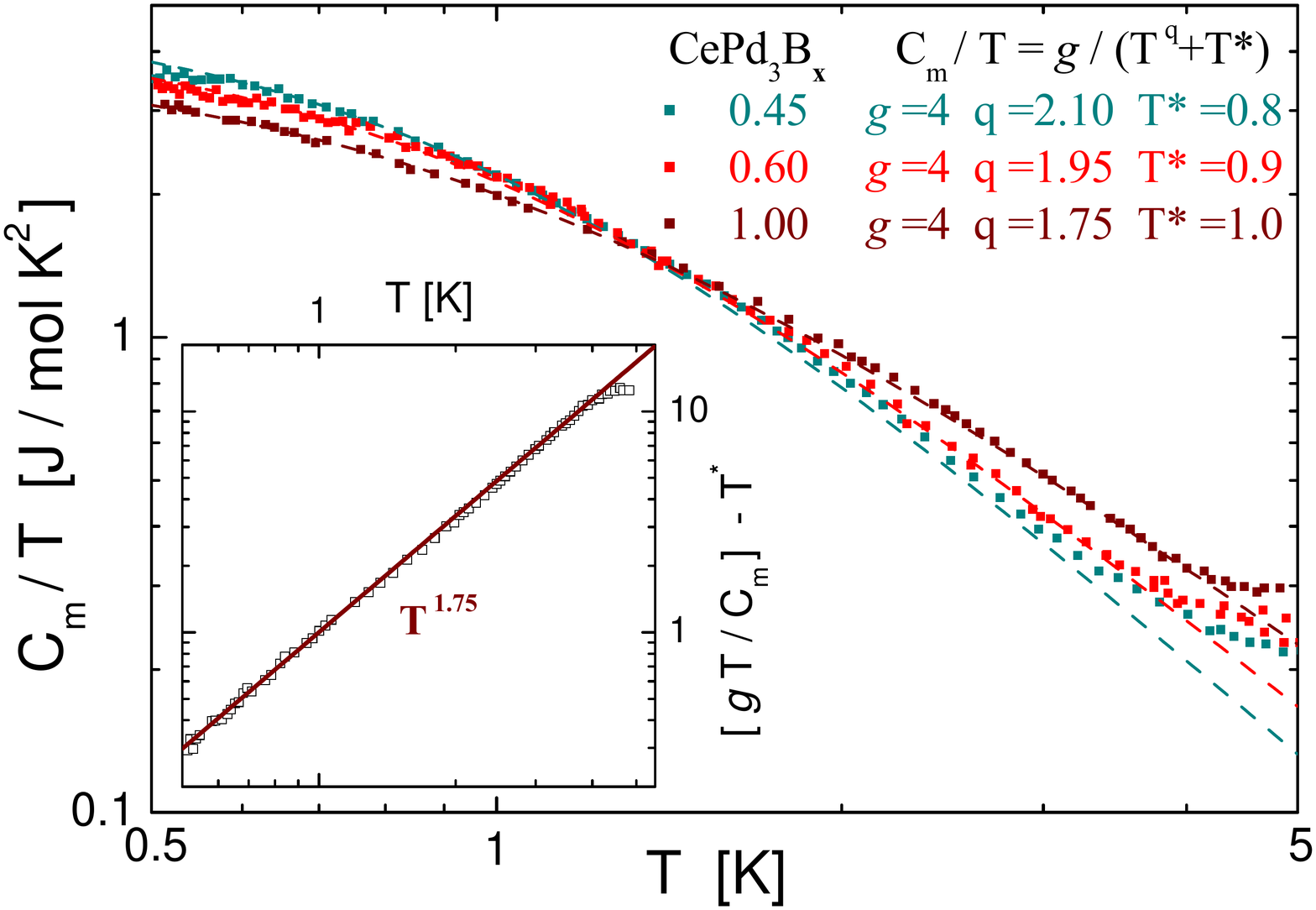}
\caption{%
(Color online) Fits for low temperature specific heat of
CePd$_3$B$_x$ compounds with a modified power law function. Inset:
normalized inverse of $C_m/T$ to extract the power law exponent
for CePd$_3$B in a double logarithmic representation (see the
text).} \label{F3}
\end{figure}

Unlike $C_m/T \propto -Ln(T/T_0)$ dependence of
CeCu$_{5.8}$Au$_{0.2}$, it is evident from Fig.~\ref{F2} that
CeNi$_9$Ge$_4$ and CePd$_3$B$_x$ compounds obey a power law al low
temperature. In order to check that behavior, we analyze in
Fig.~\ref{F3} such dependence for CePd$_3$B$_x$ compounds using a
modified power law $C_m/T=g/(T^q+T^*)$ \cite{anivHvL}, obtaining
similar values of the exponent $q=1.95\pm 0.2$ and $T^*$ values
around 1\,K for all of them. The extrapolation down to $T = 0$ is
given by $C_m/T \mid_{lim T\to 0} = g/T^*= 4.5\pm
0.5$\,J/molK$^2$. The accuracy of this description is evidenced in
the inset of Fig.~\ref{F3} where a double logarithmic
representation allows to extract the exponent $q=1.78$ for
CePd$_3$B. Notice that the applied power law excludes spin or
cluster glass behaviors because those cases follow a $C_m \propto
1/T^2$ dependence (i.e. $C_m/T \propto 1/T^3$) \cite{SpecHeat}.

\begin{figure}[t]%
\includegraphics*[width=0.6\linewidth]{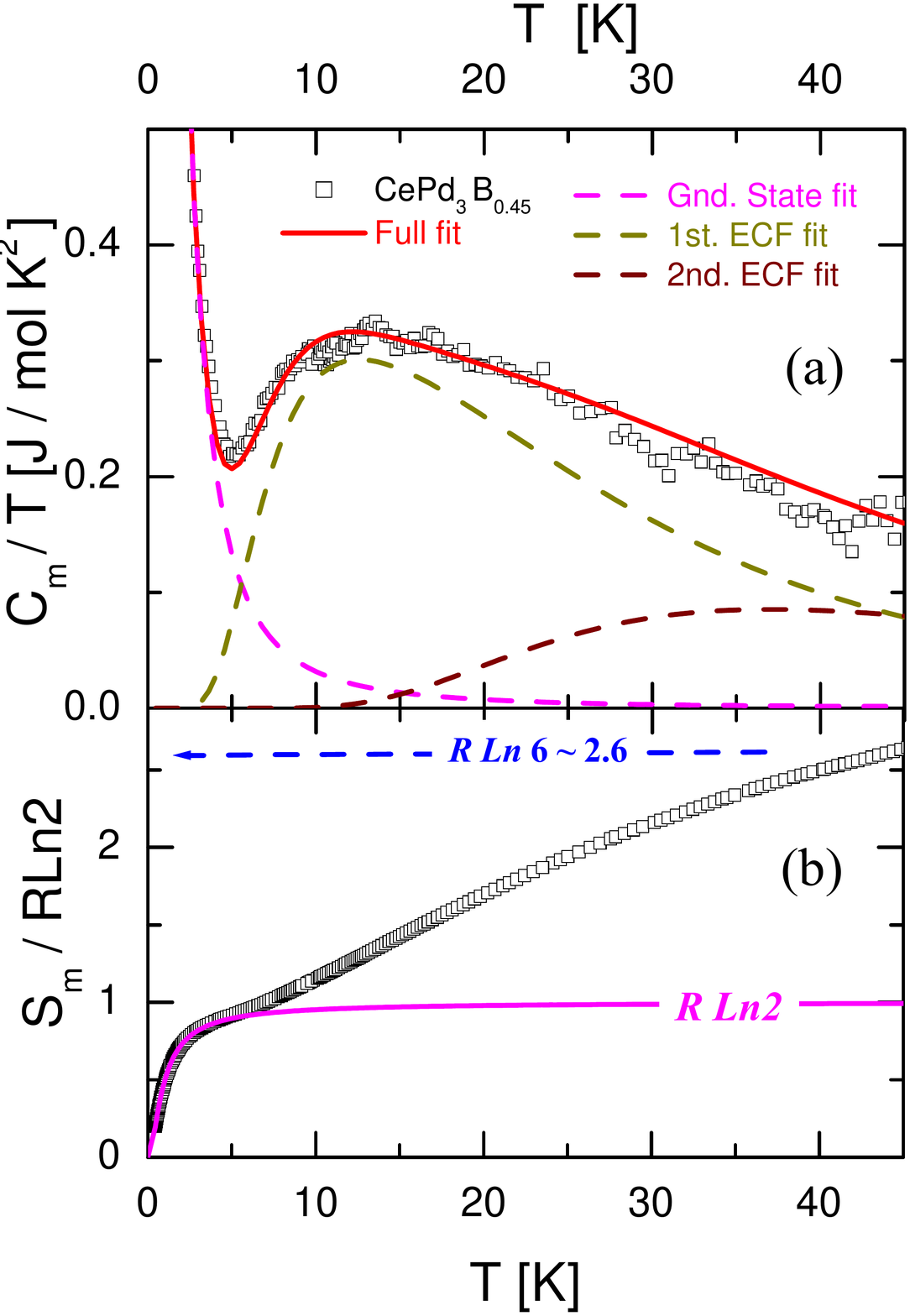}
\caption{%
(Color on line) (a) High temperature specific heat of
CePd$_3$B$_{0.45}$ described accounting for the 1st. and 2nd. ECF
doublets. (b) Thermal dependence of the magnetic entropy in R$Ln$2
units.} \label{F4}
\end{figure}

The $N=2$ GS character of cubic CePd$_3$B$_x$ compounds is
confirmed by high temperature $C_m$ measurements. Above $T\approx
5$\,K, the $C_m(T)/T$ results deviate from power law $T$
dependencies because the first ECF level starts to contribute. In
Fig.~\ref{F4}a we analyze that contribution on the high
temperature $C_m(T)/T$ results from CePd$_3$B$_{0.45}$. A good
overall fit is obtained accounting for the GS power law dependence
and two Schottky anomalies originated in the contribution of two
ECF doublets centered at $\Delta_I\approx =27$,K and
$\Delta_{II}\approx = 60$\,K respectively. In this case, a level
width corresponding to a $T_K^I$ of $\approx 10$\,K is required
for the first ECF doublet. Nevertheless, in comparison with
CeNi$_9$Ge$_4$, in this compound the $\Delta_I /T_K^I \approx 3$
ratio makes the GS to retain its actual $N=2$ character. Such a
difference is also reflected in the thermal dependence of the
entropy (see Fig.~\ref{F4}b) which shows a tendency to a $S_m \to$
R$Ln$2 saturation around 7\,K, prior the onset of the ECF levels
contribution.

\section{Discussion}

The basic thermodynamic implications of the $T\to 0$ physics
regard the third law of thermodynamics which governs the
$\lim_{T\to 0}\partial S/\partial T = C_m/T$. The conventional
scenario for $\partial S/\partial T =0$ corresponds to the case of
a singlet, realized e.g. in a long-range-order GS associated to a
positive curvature of $S_m(T)$ (i.e. $\partial^2S_m /\partial
T^2>0$) \cite{Abriata}. The case of $\partial S/\partial T \neq 0$
(with $\partial^2S_m /\partial T^2=0$) has its simplest example in
metallic systems whose conduction electrons are described as a
Fermi gas. In this case, $\partial S/\partial T = \gamma$ defines
the Sommerfeld coefficient typically in the range of a few
[mJ/molK$^2$]. Non magnetic Ce compounds in the intermediate
valence regime, behave as Fermi liquids (FL) with $10 \leq \gamma
\leq 100$\,mJ/molK$^2$ values, whereas those with $100 \leq \gamma
\leq 1$\,J/molK$^2$ are recognized as HF. Within this group, one
should distinguish between those showing FL and NFL character. The
former keep their fermionic nature and form narrow bands of heavy
quasi-particles of enhanced effective mass, typically exhibiting
$\gamma$ values up to 400\,mJ/molK$^2$. The later present a dense
spectrum of low energy excitations which not necessarily form a
band. The $C_m(T)/T$ dependence typically obeys logarithmic or
$a-b \surd T$ functions \cite{Stewart}, in coincidence with am
eventual Kondo-breakdown scenario \cite{Kbreak}. Consequently,
$C_m/T$ cannot be identified with the canonical Sommerfeld
$\gamma$ coefficient because its temperature dependence differs
from that of a FL since $C_m/T \mid_{lim T\to 0 } = \gamma \ast [1
- (T/T_K)^2]$ in the later \cite{CRC}.

\subsection{$T \to 0$ divergencies}

Even higher values are observed in a few very heavy fermion (VHF)
ranging within $1 \leq C_m/T \mid_{lim T\to 0 } < 8$\,J/molK$^2$,
which do not show magnetic order down to the mK region of
temperature. The analysis of their thermodynamic properties as $T
\to 0$ is the main scope of this work. Typical thermal
dependencies for these NFL systems are $-Ln(T/T_0)$ or $T^{-q}$
\cite{HvL,Contin,Miranda}, with the former not exceeding $\approx
3$\,J/molK$^2$ \cite{HvL1999}. Both dependencies imply that $C_m/T
\mid_{lim T\to 0}$ keeps increasing because the density of low
lying energy excitations grow continuously.

Since thermodynamic postulates and experimental evidences indicate
that singularities at $T= 0$ and consequent negative $S_m(T\to 0)$
curvature ($\partial^2S_m /\partial T^2 < 0$) are not physical,
the question arises whether there is any upper limit for the
$\partial S/\partial T$= $C_m/T$ slope. The experimental results
collected in Fig.~\ref{F2} for VHF Ce-lattice intermetallics
suggest the existence of such upper limit, reached following the
modified power law $C_m/T=g/(T^q+T^*)$.

Comparing characteristic energies of Ce compounds whose
magnetically ordered phase fulfils the condition of $S_{MO} \to 0$
when $T_{ord} \to 0$ (described as case {\bf 'I'} in ref.
\cite{PhilMag}), one finds that there is a significant coincidence
in the scales of energy extracted from different types of
analysis. In fact, the compounds fulfilling the mentioned
condition show a clear change of regime at $T^{CR} = 2.5\pm
0.3$\,K \cite{PhilMag}, whereas in those systems which do not
order magnetically the characteristic energy scale related to the
deviation from a pure power law (i.e. the parameter $T^*$
extracted from Fig.~\ref{F3} fits) is around 1\,K. If $T^{CR}$
represents a threshold between thermal and quantum fluctuations
dominated regions, the $T^*$ value indicate that the deviation
from the power law dependence may occur as an alternative to
magnetic order.

\begin{figure}[t]%
\includegraphics*[width=0.6\linewidth]{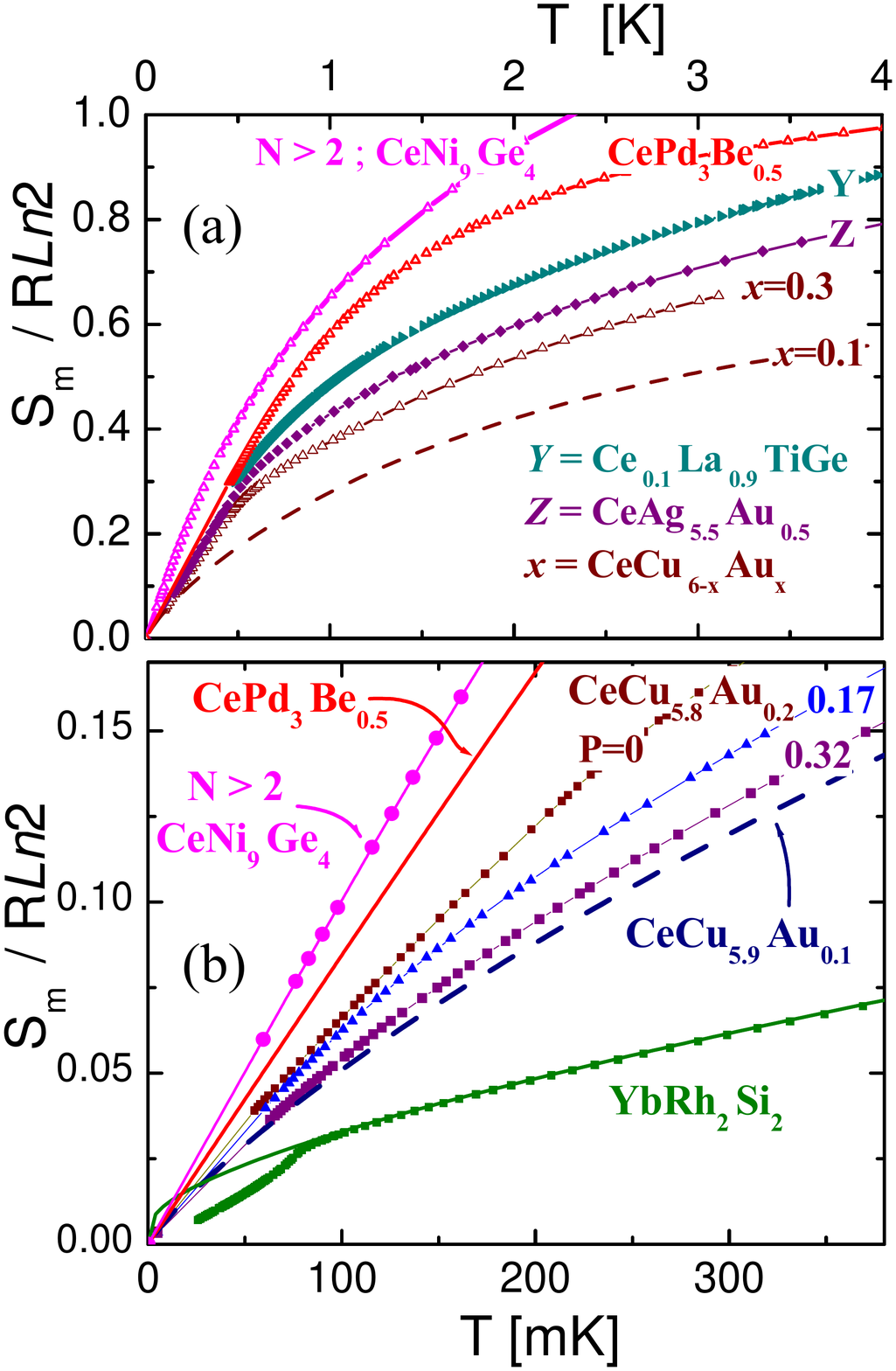}
\caption{%
(Color online) Examples of maximum slope of $S_m(T)$ for Cerium
N=2 GS systems, (a) within the  $0<T<4$\,K range including
CePd$_3$B$_{0.45}$, CeCu$_{6-x}$Au$_x$ \cite{Lohenysen},
(Ce$_{0.1}$La$_{0.9}$)TiGe \cite{CeLaTiGe}, CeAg$_{6-z}$Au$_z$
\cite{ScheidtAg}, and CeNi$_9$Ge$_4$ with $N_{eff}>2$
\cite{ScheidtGe4}. (b) Detail of the $S_m(T)$ dependence within
the mK range including: CeCu$_{5.9}$Au$_{0.1}$ (dashed line),
CeCu$_{5.8}$Au$_{0.2}$ at different pressures (in GPa) and, for
comparison, the power law fit of CePd$_3$Be$_{0.5}$ extrapolated
from $T> 0.5$\,K. The $N_{eff} > 2$ CeNi$_9$Ge$_4$ (after
\cite{ScheidtGe4}) and the entropy evaluation from the well known
YbRh$_2$Si$_2$ \cite{YbRh2Si2} are also included for a wider
comparison.} \label{F5}
\end{figure}

The question arises concerning the reason why the Ce compounds
collected in Fig.~\ref{F3} do not order magnetically down to such
a low temperature despite their robust Ce$^{3+}$ moment. A common
feature of these compounds is that the magnetic (Ce) lattice
presents a subtle displacements from their strict periodic atomic
positions frustrating the development of any long-range-order
parameter. This possibility is supported by the large available
volume of Ce atoms in CeNi$_9$Ge$_4$ and the effect of random
distribution of $B$ or $Be$ interstitials in CePd$_3$X$_x$.
Preliminary studies on a novel compound with the TiNiSi type
structure showing very similar $C_m(T)$ dependence \cite{Isolde}
confirms these observations.

As remarked in Ref.\cite{Kbreak}, in HF systems tuned by chemical
pressure the coupling of electrons to lattice degrees of freedom
can play a nontrivial role. In our case, due to the very low value
of the characteristic energy scale, the local moments may form a
spin-liquid state dominated by quantum fluctuations \cite{Kbreak}
without braking any symmetry (i.e. no phase transition is
required). In fact, electrical resistivity of CePd$_3$B
continuously increases at low temperature not showing any
coherence effect \cite{Gladys}, see the inset in Fig.~\ref{F2}.
This scenario does not map a single impurity picture because, like
La doped systems (Ce,La)Ni$_9$Ge$_4$ \cite{ScheidtGe4},
(Ce$_{0.1}$La$_{0.9})$TiGe \cite{CeLaTiGe} and
(Ce$_{0.03}$La$_{0.97})$B$_6$ \cite{satoB6} which have quite high
$C_m/T$ values, they do not show evidences for $C_m/T \mid_{lim
T\to 0}$ saturation but a $-Ln(T/T_0)$ type dependence. Notably,
the power law temperature dependence of Ce-lattice CeNi$_9$Ge$_4$
transforms into a $-Ln(T/T_0)$ one by La doping in
(Ce,La)Ni$_9$Ge$_4$ \cite{ScheidtGe4}. The universality of a
logarithmic type of temperature dependence is remarked by the
existence of a single expression $C_m/t = - 7.2 Ln (t)$
\cite{scaling} once the temperature is normalized by $t = T/T_0$,
being $T_0$ a characteristic temperature of each system.

\subsection{Entropy}

In this subsection we will discuss the thermodynamic consequences
of a $C_m/T = \partial S/\partial T $ upper limit on the thermal
variation of the entropy at the $T\to 0$ limit. In Fig.~\ref{F5}a
we have collected the low temperature $S_m(T)$ dependencies
extracted from a number of $N=2$ VHF Ce systems reaching $C_m/T
\geq 3$\,J/molK$^2$ values, independently of their temperature
dependencies and whether they order or not. That figure includes
CePd$_3$Be$_{0.5}$, two exemplary concentrations of the well known
CeCu$_{6-x}$Au$_x$ \cite{Lohenysen} and CeCu$_{5.5}$Ag$_{0.5}$
\cite{ScheidtAg}. Also diluted (Ce$_{0.1}$La$_{0.9})$TiGe
\cite{CeLaTiGe} is included to confirm that this limit is
independent of lattice configurations, and the entropy of
CeNi$_9$Ge$_4$ \cite{ScheidtGe4} for comparison with a $N_{eff} >
2$ compound.

As it can be seen, experimental results suggest a sort of {\it
entropy envelope curve} for $N=2$ GS systems which is
qualitatively represented by the $S_m(T)$ dependence of
CePd$_3$Be$_{0.5}$. The highest $\partial S/\partial T$ is shown
by the $N_{eff} >2$ GS compound CeNi$_9$Ge$_4$, which is overcome
by Ce-diluted (Ce$_{0.03}$La$_{0.97})$B$_6$ \cite{satoB6} (not
shown) which derives from the $N=4$ GS compound CeB$_6$ with BCC
structure. A closer analysis of the mK region is presented in
Fig.~\ref{F5}b, with the {\it entropy envelope curve} of
CePd$_3$B$_{0.45}$ extrapolated from the $T>0.5$\,K fit which
shows a $C_m/T \mid_{lim T\to 0} \approx 5$\,J/molK$^2$. Notably,
Ce-lattice systems claimed to diverge as $-Ln(T/T_0)$ fit into
that limit. In fact, the highest HF system CeCu$_{6-x}$Au$_x$
shows a $C_m/T = -0.63 Ln(T/5.27)$ dependence \cite{Lohenysen}
which, at the extremely low temperature magnetic order of $T_N
\approx 2$\,mK \cite{onuki}, reaches the expected limit value of
$\approx 5$\,J/molK$^2$.

For comparison with other Rare earth compounds, the thermal
dependence of the entropy of YbRh$_2$Si$_2$ is included in
Fig.~\ref{F5}b. Also in this case a first order transition occurs,
at $\approx 80$\,mK \cite{YbRh2Si2}, close to the {\it entropy
envelope curve}, whereas the curve extrapolated from the
paramagnetic state using the modified power low function would
have exceed that value below about 40\,mK. Similar situation
occurs with the novel Yb compound YbNi$_4$P$_2$ \cite{Krellner}
which also undergoes a first order transition at $T_C = 0.17$\,K
approaching the same {\it envelope curve}. In both cases $C_m/T
\mid_{lim T\to 0}\approx 2$\,J/molK$^2$, which is much smaller
than 5\,J/molK$^2$ reported for YbCu$_{5-x}$Au$_x$ \cite{Galli}
and the record one for YbCo$_2$Zn$_{20}$ of 7.8\,J/molK$^2$
\cite{Tori}. This compound does not order magnetically down to the
mK range of temperature, but is considered to have a N = 4 GS
\cite{Tori}.

\section{Summary}

Using selected specific heat results, we have analyzed the
thermodynamic implications of the $T\to 0$ physics in VHF Ce
compounds which do not order magnetically but exhibit extremely
high $C_m/T \mid_{lim T\to 0}$ values. From these experimental
information we have observed that: i) according to thermodynamic
laws, $C_m(T)/T$ power law temperature dependencies tend to
saturate, ii) the onset of that saturation occurs within the range
of temperature dominated by quantum fluctuation (i.e. $T<2$\,K),
which may lead to access to an exotic GS like e.g. a spin liquid
one. iii) Alternatively to saturation the systems undergo magnetic
transitions because of the entropy accumulation at very low
temperature. iv) An empirical upper limit of $C_m/T\mid_{\lim T\to
0} \approx 4.5$\,J/molK$^2$ for Ce compounds with $N=2$ GS is
observed. Higher values are found in compounds with $N_{eff} >2$
ground state. These experimental observations highlight the role
of thermodynamic laws in the understanding of realistic $T\to 0$
physics, being specific heat and entropy the tools to distinguish
between real and 'wished' candidates to quantum critical regime.

We conclude that thermodynamic laws and quantum critical
mechanisms intervene simultaneously in the GS formation. Third law
constraint on entropy accumulation at $T \to 0$ interdicts $C_m/T$
singularities imposing an upper limit to the density of low lying
quantum excitations. This constraint drives the systems to
alternative GS through first order transitions (even at a few mK)
or to the formation of exotic states. The access to a quantum
critical point a $T \to 0$ seems to be limited by simple
thermodynamic conditions before a technical limit of the cooling
process occurs.

\subsection*{Acknowledgements}
The author acknowledges E. Bauer, J.P. Kappler, G. Nieva, T. Radu,
E-W. Scheidt, G. Schmerber and I. Zeiringer for allowing to access
to original experimental data.


\begin{thebibliography}{99}

\bibitem{HvL} H.v. L\"ohneysen, A. Rosch, M. Vojta, P. W\"olfle; Rev. Mod. Phys. 79
(2007) 1015.

\bibitem{PhilMag} J.G. Sereni, Philosophical Magazine, 'overview' iFirst, 11 Sept 2012, 1-25. \& ArXiv:
cond-mat.1202.1724, 8 Feb 2012.

\bibitem{TVojta} T. Vojta; Ann. Phys. (Leipzig) 9 (2000) 403.

\bibitem{Pippard} See for example: A. B. Pippard, in {\it Elements of Classical Thermodynamics for Advanced
Students of Physics}, Cambridge University Press, 1957.

\bibitem{anivHvL} J.G. Sereni; J. Low Temp. Phys. 147 (2007).
179.

\bibitem{CePd3B} J.G. Sereni, G. Nieva, J. Kappler, M. Besnus, A. Meyer, J. Physics F (Metal Phys.) 16 (1986) 435.

\bibitem{Gladys} G. Nieva, PhD Thesis, {\it Effects of chemical pressure produced by atomic interstitial inclussion into CePd$_3$
}, University of Cuyo, Argentina, 1988.

\bibitem{ScheidtGe4} U. Killer, E-W. Scheidt, W. Scherer, H. Michor, J. Sereni, Th. Pruschke and S. Kehrein, Phys. Rev. Lett. 93 (2004) 216404.

\bibitem{HvL1999} H.v. L\"ohneysen; J. Magn. and Magn. Mater. 200
(1999) 532.

\bibitem{Handb} J.G. Sereni, in {\it Hanbook en  Handbook on the Physics and Chemistry of Rare Earths},
Eds.: K.A. Gschneidner Jr. and L. Eyring, 1991, Vol. 15, ch. 98,
Elsevier Science Pub. B.V.

\bibitem{SpecHeat} J.G. Sereni, in {\it Encyclopaedia of Materials: Science and Technology}. Eds: K.H. Buschow and E. Gratz, Elsevier Science, 2001

\bibitem{Abriata} J.P. Abriata and D.E. Laughlin, Prog. Matterial
Science 49 (2004) 367.

\bibitem{Kbreak} A. Hackl and M. Vojta; Phys. Rev. B 77 (2008) 134439.

\bibitem{scaling} J.G. Sereni, C. Geibel, M. G\'omez Berisso, P. Hellmann, O. Trovarelli and  F. Steglich,
Physica B 230 (1997) 580.

\bibitem{Stewart} G.R. Stewart, Rev. Mod. Phys. 73 (2001) 797.

\bibitem{CRC} See for example N.E. Phillips, in {\it CRC Critical
Reviews in Solid State Sciences}, Eds. D. Schuele and R.W.
Hoffman, Vol 2, p. 467, 1971.

\bibitem{Contin} M.A. Continentino, Brazilian Jour. of Phys. 41 (2011) 201.

\bibitem{Miranda} E. Miranda and V. Dobrosavljevi\'c; Rep. Prog.
Phys. 68 (2005) 2337.

\bibitem{Desgr} See for example H.-U. Desgranges and K.D. Schotte, Physics Letters
91A (1982) 240.

\bibitem{Isolde} I. Zeiringer, Institute of Physical Chemistry, University of Vienna, private communication (2012).

\bibitem{CeLaTiGe} J. G Sereni, M. G.-Berisso, M. Deppe, N.C.Canales, C. Geibel; Phys. Stat. Solidi (b), 247 (2010) 707.

\bibitem{satoB6} N. Sato, M. Takahashi, T. Kashima, K. Sugiyama,
M. Date, T. Satoh and T. Kasuya; J. Magn. and Magn. Mat. 52 (1985)
250.

\bibitem{Lohenysen} H.v. L\"ohenysen, T. Pietrus, G. Portisch, A.
Schr\"oder, H.G. Schlager, M. Sieck, T. Trappmann, Phys. Rev.
Lett. 72 (1994) 3262.

\bibitem{ScheidtAg} K. Heuser, E.-W. Scheidt, T. Schreiner, G.R. Stewart
Phys. Rev. B Rapid Comm. 58, R15959, (1998)

\bibitem{onuki}H. Tsujii, E. Tanaka, Y. Ode, T. Katoh, T. Mamiya,
S Arai, R. Settai, Y. \~Onuki; Phys. Rev. Lett. 84 (2000) 5407.

\bibitem{YbRh2Si2} C. Krellner, S. Hartmann, A. Pikul, N. Oeschler, J. G. Donath, C. Geibel, F. Steglich, J. Wosnitza;
Phys. Rev. Lett. 102 (2009) 196402.

\bibitem{Krellner} C. Krellner, S. Lausberg, A Steppke, M. Brando, L. Padrero, H. Pfau, S. Tenc\'e, H. Rosner, F. Steglich, C. Geibel;
New Jour. of Physics 13 (2011) 103014.

\bibitem{Galli} M. Galli, E. Bauer, St. Berger, Ch. Dusek, M. Della Mea, H. Michor, D. Kaczorowski, E.W. Scheidt, F. Mirabelli; Physica B 312-313
(2002)489.

\bibitem{Tori} M.S. Torikachvili, S. Jia, E.D. Mun, S.T. Hannahs, R.C. Black, W.K. Neils, D. Martien, S.L. Bud'ko,
P.C. Canfield; PNAS 104 (2007) 9960.

\end{thebibliography}
\end{document}